\documentclass[aps,prl,showpacs,twocolumn,amsmath,amssymb]{revtex4} 
\usepackage[english]{babel}
\usepackage{latexsym}
\usepackage{graphics}
\usepackage{subfigure}
\usepackage{epsfig}
\begin{document}

\title{Atomic Fermi-Bose Mixtures in Inhomogeneous and Random Lattices:\\ 
 From Fermi Glass to Quantum Spin Glass and Quantum Percolation}
\author{A. Sanpera$^{1,2}$, A. Kantian$^2$, 
L. Sanchez-Palencia$^2$, J. Zakrzewski$^3$, and  M. Lewenstein$^{1,2}$} 
 
\affiliation{$^1$Institut de Ci\`encies Fot\`oniques, 
08034 Barcelona, Spain\\
$^2$Institut f\"ur Theoretische Physik, Universit\"at Hannover, 
D-30167 Hannover,Germany\\
$^3$Instytut Fizyki,
  Uniwersytet Jagiello\'nski, PL-30-059 Krak\'ow, Poland}

\date{\today}

\pacs{03.75.Kk,03.75.Lm,05.30.Jp,64.60.Cn}

\begin{abstract}
We investigate atomic Fermi-Bose mixtures in inhomogeneous and random 
optical lattices in the limit of strong atom-atom interactions. 
We derive the effective Hamiltonian describing the dynamics of the system 
and discuss its low temperature physics. 
We demonstrate possibility of controlling the 
interactions at local level in inhomogeneous but regular lattices. 
Such a control leads to the achievement of Fermi glass, 
quantum Fermi spin glass, and quantum percolation regimes 
involving bare and/or composite fermions in random lattices.  
\end{abstract}

\maketitle


Fermi-Bose (FB) mixtures attract considerable interest 
in the physics of ultra-cold atomic and molecular gases,  comparable with 
the interest in  molecular Bose-Einstein condensation (BEC) \cite{molec}, or 
Bardeen-Cooper-Schrieffer (BCS) transition \cite{revBCS} in ultra-cold 
Fermi mixtures.
The reason for interest in FB systems is threefold. 
First, these are very fundamental systems that have no direct analogues 
in condensed matter physics. Second, these systems can be very 
efficiently cooled using  sympathetic cooling
down to very low temperatures $T$ (of order of tens nK)
\cite{bfcooling1,bfcooling2,bfcooling3,bfcooling4}. Finally, their physics is 
extremely rich and not yet fully understood.

FB mixtures have been intensively
 studied in traps \cite{bulk}, but 
the experimental observation of  the superfluid (SF) to 
Mott-insulator (MI) transition in bosonic gases \cite{bloch}, 
predicted in Ref. \cite{jaksch}, has 
triggered the interest in the physics of FB mixtures in optical 
lattices \cite{modugnoetal}. 
Under appropriate conditions such mixtures can be well described 
by the Fermi-Bose Hubbard model (FBH) \cite{eisert}. 
A particularly appealing feature of the FBH model
is the possibility to produce novel quantum phases 
\cite{BlatterRothBurnett}, fermion-boson induced superfluidity \cite{Illu}, 
and composite fermions, which for attractive (repulsive) interactions 
between fermions and bosons, are formed by a fermion and 
bosons (bosonic holes) as first shown in \cite{svistunov} (see also 
\cite{lewen,kagan}). 

FB mixtures in the limit of strong atom-atom interactions (strong 
coupling regime) show a very rich 
variety of quantum phases in periodic optical lattices \cite{lewen}. 
They include 
the mentioned composite fermions, and range from a normal Fermi liquid, 
a density wave, a superfluid liquid, to an insulator with fermionic domains. 
The phase diagram of the system has been determined 
in Ref.~\cite{fbmean} by means of mean-field theory 
\cite{sachdev}. These studies have been generalized 
recently to inhomogeneous lattices \cite{neweisert} to include 
the effects of the lattice and of a possible trap potential. 
So far, only the case of strong interactions and vanishing 
hopping has been considered.

In the present Letter we study the low temperature physics of FB mixtures
in optical 
lattices with local and random inhomogeneities in the strong 
interactions limit but including tunneling as a perturbation. 
We  show that interactions and tunneling may be controlled 
at the local level in inhomogeneous lattices \cite{duan}. 
This control gives access to a wide variety of regimes and we derive 
the corresponding effective Hamiltonians. 
We then show how to achieve Fermi glass, fermionic spin glass and 
quantum percolation regimes involving bare and/or composite 
fermions in random lattices.

We consider a sample of ultra-cold bosonic and 
(polarized) fermionic atoms (e.g.
$^7$Li-$^6$Li or $^{87}$Rb-$^{40}$K) trapped in an optical lattice. 
At low temperature, the atoms occupy only the lowest energy band and 
it is convenient to work in the corresponding Wannier basis \cite{jaksch}. 
Note that a fermion number $N_\textrm{F}$ strictly smaller than the 
number of lattice sites $N$ is required here.
The Hamiltonian of the system reads \cite{eisert,Auerbach}:
\begin{eqnarray}
&&H_{\mathrm{FBH}}=-\sum_{\left\langle ij\right\rangle }
(J_\textrm{B} b_{i}^{\dagger }b_{j}+J_\textrm{F} f_{i}^{\dagger }f_{j}+ {\rm h.c.})  \\
&&+\sum_{i}\left[
\frac{1}{2}Vn_{i}(n_{i}-1)+Un_{i}m_{i}-\mu^B_i n_{i}-\mu^F_i m_i\right] ,
\nonumber 
\label{Hamiltonian} 
\end{eqnarray}
 where $b_{i}^{\dagger }$, $b_{j}$, $f_{i}^{\dagger }$, 
$f_{j}$ are the bosonic and fermionic
creation-annihilation operators,  $n_{i}=b_{i}^{\dagger
}b_{i}$ and $m_{i}=f_{i}^{\dagger }f_{i}$.
The FBH model describes: i) nearest neighbor (n.n.) 
boson (fermion) hopping, with an  
associated negative energy, $-J_\textrm{B}$ ($-J_\textrm{F}$); 
ii) on--site repulsive boson-boson interactions with
an energy $V$; iii) on--site boson-fermion interactions with an
energy $U$, which is positive (negative) for repulsive
(attractive) interactions, iv) and, finally, interactions with the optical 
potential, with energies $\mu^\textrm{B}_i$ and $\mu^\textrm{F}_i$. 
In the following, we shall consider only the case $J_\textrm{B}=J_\textrm{F}=J$
and the regime of strong interactions, namely $V,U\gg J$. 

In a periodic optical lattice, $\mu^\textrm{B,F}_i$ is simply the (bosons or 
fermions) chemical potential and is independent of the site $i$. 
It is, however, possible to add a laser field independent of the lattice to 
modify the depths of the optical potential wells in a site-dependent way
\cite{boseglass}. 
In this case, the local potential depth has to be added to 
$\mu^\textrm{B,F}_i$, which may now be inhomogeneous.
If the added field is periodic and if the spatial period is commensurate 
with the lattice period, $\mu^\textrm{B,F}_i$ is periodic; if the spatial
periods are incommensurate, $\mu^\textrm{B,F}_i$ is quasi-periodic. 
One can also add a random 
speckle field, so that $\mu^\textrm{B,F}_i$ is random. Experimental 
techniques offer full possibilities to control such periodic, quasi-periodic,
or disordered $\mu^\textrm{B,F}_i$ \cite{grynberg}. Note,
that the additional inhomogeneous potential might, but does not 
have to, act equally on both atomic species. Here, we will study the case 
$\mu^F_i=0, \mu^B_i=\mu_iV$.


 In Ref.~\cite{lewen}  we have used 
the method of degenerate second order perturbation theory to 
derive an effective Hamiltonian by projecting the wave function onto 
the multiply 
degenerate ground state of the system in the absence of tunneling. 
This can be extended to the present situation, where
there are 
very many states with similar energies. It is thus reasonable 
to project the wave function on the manifold of `ground states'. 
These states are local minima 
of energy, since  at least some of hopping acts increase their 
energy by $V$ or $|U|$. 

Let us consider first $J=0$, and  
the case $0\le \mu_i<1$.
In the absence of a fermion one 
expects one boson per site, i.e. $n_i=1$ 
\cite{fillingfactor}.
We shall consider here only the case of 
repulsive interactions, i.e. $\alpha =U/V> 0$.  

It is useful to divide the sites into : i) $A$--sites, 
for which $\mu_i-\alpha\ge 0$, and fermions do not push bosons out,  and
ii) $B$--sites, for which $\mu_i-\alpha < 0$, and the fermion pushes the
boson out forming a composite fermion-bosonic hole. 
Energetically, the second situation
is favorable, so for a given set of $N^\textrm{A}$  of $A$-, and $N^\textrm{B}=N-N^\textrm{A}$ 
of  $B$--sites,
the fermions at very low temperature will first occupy the $B$--sites 
until $N_\textrm{F}=N^\textrm{B}$, and then they will start to occupy the $A$--sites. 
We construct the corresponding projector operators $P$, $Q=1-P$, 
which depend on $N^\textrm{A}$ and $N_\textrm{F}$. The  operator $P$ describes 
the
projection onto the manifold of quasi-degenerated states 
in which the fermions occupy the $B$--sites stripped of bosons and  
some $A$--sites only if $N_\textrm{F}\ge N^\textrm{B}$. In this case
there is a boson in any $A$--site and if  $N_\textrm{F}< N^\textrm{B}$ 
there is also a boson in the $B$--sites which do not contain fermions.
We use the formalism of second order time dependent perturbation theory
\cite{cohen}, and project the Schr\"odinger equation,
$i\hbar \partial_t|\psi(t)\rangle=(H_0+H_1)|\psi(t)\rangle,$
onto the manifold of states spanning $P$. The ``zeroth-order'' part $H_0$  
contains the atomic interactions
and terms proportional to the chemical potential and  
commutes with $P$. $H_1$ represents the tunneling terms.
The effective equation for $|\psi_P\rangle=P|\psi\rangle$ reads then
$i\hbar\partial_t|\psi_P(t)\rangle=H_{eff}|\psi_P(t)\rangle$
where
\begin{eqnarray}
&&\langle out | H_\textrm{eff}|in\rangle =
\langle out |H_{0}|in\rangle \label{hamil} 
+\langle out|PH_{1}P|in\rangle\\
&&-\frac{1}{2}\langle out|
PH_1Q\left(\frac{1}{H_0-E_\textrm{in}}+\frac{1}{H_0-E_\textrm{out}}\right)
QH_1P|in\rangle.\nonumber
\end{eqnarray}
The effective Hamiltonian $H_{eff}$ has the form
\begin{equation}
H_\textrm{eff}=\sum_{\left\langle ij\right\rangle}[-(d_{ij}F_{i}^{\dagger
}F_{j}+h.c.)+K_{ij}M_{i}M_{j}]+\sum_i \tilde\mu_iM_i,
\label{EffHamiltonian}
\end{equation}
where $F_i,F_i^{\dag}$ are the 
corresponding (composite) fermionic annihilation and 
creation operators, and $M_i=F_{i}^{\dagger
}F_{i}$. 
The hopping amplitudes $d_{ij}$ and the n.n. couplings $K_{ij}$ (which might
be repulsive ($>0$) or attractive ($<0$)) are of 
order of $J^2/V$. The values of the couplings  depend on  
$\alpha$, $\tilde\mu_{i}$, $\tilde\mu_{j}$, 
and $J$, and have to be determined carefully for different cases, as 
discussed below. 
Note, however, that  
the  hopping $i\to j$, or back causes the energy change 
$\pm (\Delta_{ij}=\mu_i-\mu_j)$ in units of $V$, 
i.e is highly non-resonant and inefficient for  
$\Delta_{ij}\simeq 1$; it will first lead  to jump rates of 
order 
$O(J^4/V^3)$. 
 Additionally,
composite fermions may feel the local 
energy $\tilde \mu_i$.

I) {\it All sites are of type B}. In this case we have a gas of composites 
flowing within  the MI 
with 1 boson per site. The couplings are
\begin{eqnarray}
d_{ij}&=&\frac{J^2}{V}\left(\frac{\alpha}{\alpha^2-(\Delta_{ij})^2}
+\frac{1}{\alpha}\right)\label{dij}\\
K_{ij}&=&-\frac{J^2}{V}\left(\frac{4}{1-(\Delta_{ij})^2}-\frac{2}{\alpha}-
\frac{2\alpha}{\alpha^2-(\Delta_{ij})^2}\right).\label{kij}
\end{eqnarray}
The chemical potential $\tilde\mu_i/V\simeq\mu_i $ up to corrections 
of order $O(J^2/V)$. The hopping amplitudes $d_{ij}$ are for this case 
always positive, although may vary quite significantly with disorder, 
especially when $\Delta_{ij}\simeq\alpha$. 
As shown in Fig. 1, for $\alpha>1$,  $K_{ij}\le 0$ and we deal  with
attractive (although random) interactions. For  $\alpha<1$, but close to 1,
 $K_{ij}$ might take positive or negative values for  $\Delta_{ij}$ small or 
$\Delta_{ij}\simeq \alpha$. In this case {\it the qualitative character 
of interactions is controlled by inhomogeneity}.

\begin{figure}[ht] 
\begin{center}
\psfig{file=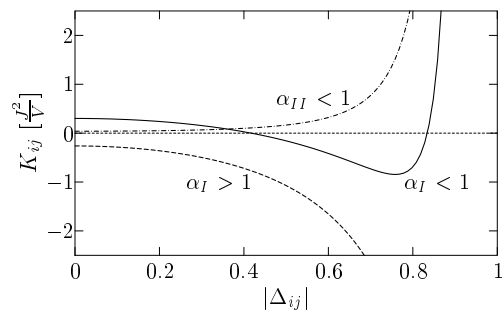,width=6.5cm}
\end{center} 
\caption{Nearest neighbor couplings $K_{ij}$  as
a function of $\Delta_{ij}$. Solid line: Coupling in case I, 
with $\alpha_{I}=0.93$. 
Dashed line: Same expression with $\alpha_{I}=1.07$. 
Dashed dotted line: Coupling in case II with $\alpha_{II}=0.03$. 
}
\label{fig:1}  
\end{figure}

At low temperature the physics of the system depends on the 
relation between  $\mu_i$'s and $\alpha$. For small inhomogeneities, 
we may neglect the contributions of $\Delta_{ij}$ to  $d_{ij}\simeq d$ 
and $K_{ij}\simeq K$, and keep only the leading disorder contribution in 
$\tilde \mu_i$. Note, that the latter contribution is relevant 
in 1D and 2D leading to Anderson localization 
of single particles \cite{gang}.   
When $K \ll d$ the system will then be in the Fermi glass phase, 
i.e. Anderson localized (and many-body corrected) single particle 
states will be occupied according to the Fermi-Dirac 
rules \cite{fermiglass}.  For repulsive interactions and $K\gg d$, 
the ground state will be a Mott insulator for large enough filling factors.
In particular, for  
filling factor 1/2 one expects 
to form a  checker-board phase. For 
intermediate values of $K/d$ delocalized metallic phases with enhanced 
persistent currents are possible \cite{metalglass}. 
Similarly, for attractive 
interactions ($K<0$) and  $|K|< d$ one expects 
competition between  pairing of fermions and disorder. 
In this case for $|K|\gg d$, the fermions 
will form a domain insulator. 

Another interesting limit is when $|\Delta_{ij}|\simeq\alpha\simeq 1$.
The tunneling becomes then non-resonant and 
negligible, while the couplings $K_{ij}$ fluctuate strongly.
We  end up then with the (fermionic) Ising spin glass model
\cite{oppermann} described by the Edwards-Anderson model \cite{parisi}: 
\begin{equation}
H_\textrm{E-A}=\frac{1}{4}\sum_{\left\langle ij\right\rangle }
K_{ij}s_is_j +\sum_{i}\tilde\mu_i s_i/2 , 
\label{HamiltonianSG}
\end{equation}
with $s_i=2M_i-1=\pm 1$. The above Hamiltonian 
is well approximated  
by a random one with Gaussian and independent  distributions 
for 
$K_{ij}/4$ 
and $\tilde\mu_i/2$ with mean 0 ($H$), and variances $K$ 
($h$), respectively. 
In this limit the system may be 
used to study various open questions of spin glass physics, 
concerning the nature of ordering (Parisi's \cite{parisi} versus
``droplet'' picture \cite{huse,stein}), broken symmetry and dynamics 
in classical (in absence of hopping) and quantum (with small, but 
nevertheless present hopping) spin glasses \cite{sachdev,georges}.
The predictions of Parisi's
mean field theory for the  model  (\ref{HamiltonianSG}) can be obtained 
by  replacing the model
 by the corresponding Sherrington-Kirkpatrick model, and employing the 
standard method of replica trick \cite{parisi}. The calculations 
differ from the standard ones in that the constraint of fixed mean number of 
fermions is applied, and one deals
 simultaneously with random couplings and 
``magnetic fields'' $\tilde\mu_i$. Following de Almeida 
and Thouless (A-T) approach \cite{at}, we obtain the A-T surface separating 
the stable paramagnetic state from the ``true'' spin glass state, 
characterized by 
replica symmetry breaking, and ultrametrically arranged 
ground states. The paramagnetic state is stable for 
\begin{equation}
\left(\frac{k_\textrm{B}T}{K}\right)^2>\left\langle\left\langle{\rm sech}^4\left
(\frac{x\sqrt{K^2q+h^2}+H}{k_\textrm{B}T}\right)\right
\rangle\right\rangle_x,
\label{atline}
\end{equation} 
where $q=\left\langle\left\langle\tanh^2\left
(\frac{x\sqrt{K^2q+h^2}+H}{k_\textrm{B}T}\right)\right
\rangle\right\rangle_x$, the constraint is  
$m=\left\langle\left\langle\tanh\left
(\frac{x\sqrt{K^2q+h^2}+H}{k_\textrm{B}T}\right)\right
\rangle\right\rangle_x$, with $m=2N_\textrm{F}/N-1$
and $\langle\langle.\rangle\rangle_x$ denotes averaging 
over normally distributed random variable $x$ which represents disorder within
the replica method \cite{parisi}. Note, that   according 
to the predictions of the alternative ``droplet'' model \cite{huse}, 
applied to  (\ref{HamiltonianSG}), no A-T surface 
is expected to exist.

 II) {\it All sites are of type A}. In this case $\alpha<1$, and 
we have a gas 
of bare fermions flowing over the MI 
with 1 boson per site. 
The coefficients are
\begin{eqnarray}
d_{ij}&=&J,\ \ \ 
K_{ij}=-\frac{J^2}{V}\left(\frac{8}{1-(\Delta_{ij})^2}\right.\\&-&
\left.\frac{4(1+\alpha)}{(1+\alpha)^2-(\Delta_{ij})^2}-\frac{4(1-\alpha)}
{(1-\alpha)^2-(\Delta_{ij})^2}\right),\nonumber\label{kij1}
\end{eqnarray}
and $\tilde\mu_i\simeq 0$ 
up to corrections of order $O(J^2/V)$. 
 The couplings  $K_{ij}$ are positive, and for 
$\alpha\simeq 0$,  $K_{ij}\simeq O(\alpha^2)$, and both the 
repulsive interactions, and disorder are very weak, leading to 
 a Fermi liquid behavior at low $T$.  
For   finite $\alpha$, and $\Delta_{ij}\simeq 
1-\alpha$, however, the fluctuations of $K_{ij}$ might be quite large.
Note, that for  $\alpha\simeq 1$, this will occur even 
for small disorder. Assuming for simplicity that $K_{ij}$ 
take either very large, or zero value, we see that the physics of bond 
percolation \cite{percbook} will play a role. The bonds will form 
a ``weak'' and ``strong'' clusters, each of which may be percolating. 
The fermions will hope freely in the ``weak'' cluster; only one fermion 
per bond will be allowed in the ``strong'' cluster.

III) {\it Both $N^\textrm{A}$ and $N^\textrm{B}$ of order $N/2$}. In this case 
the physics of site percolation \cite{percbook} will be relevant.
If $N_\textrm{F}\le N^\textrm{B}$  the composite fermions will move 
within a cluster of $B$ sites. When $N^\textrm{B}$ is above 
the classical percolation threshold, this cluster will be percolating.
The expressions Eq. (\ref{dij}) and Eq. (\ref{kij}) will still be valid,
 except that they will connect only the $B$--sites.

The physics of the system will be similar as in the case I), 
but it will occur now
on the percolating cluster. For small disorder, 
and  $K\ll d$ the system will be in a Fermi glass phase 
in which the interplay between the   Anderson localization 
of single particles due to fluctuations of $\mu_i$ and quantum percolation
effects, that is randomness of the $B$--sites cluster,  will occur.    
For repulsive interactions and 
$K\gg d$, the ground state will be 
a Mott insulator on the cluster for large filling factors.
It is an open question whether 
the delocalized  metallic phases with enhanced 
persistent current of the kind discussed in Ref. \cite{metalglass} 
might exist in this case. Similarly, it is an open question whether 
for attractive 
interactions ($K<0$) and  $|K|< d$
pairing of (perhaps localized) fermions
will take place. 
If $|K|\gg d$, we expect the fermions 
to form a domain insulator on the cluster. 

In the ``spin-glass'' limit $\Delta_{ij}\simeq\alpha\simeq 1$, 
we deal with the Edwards-Anderson spin glass on the cluster.
Such systems are of interest in condensed matter physics 
\cite{spiperc}, and again questions connected to the 
nature of spin glass 
ordering may be studied in this case.

When $N_\textrm{F}>N^\textrm{B}$, all $B$--sites will be filled, 
and the physics will occur on the cluster of $A$ sites. 
For $\alpha\simeq 0$, we shall deal
with a gas with very weak repulsive interactions, and no significant 
disorder on the random cluster. This is an ideal test ground to study 
quantum percolation at low T.  
For   finite $\alpha$, and $\Delta_{ij}\simeq 
1-\alpha$, the interplay between the fluctuating  repulsive $K_{ij}$'s 
and quantum percolation might be studied.

Summarizing, we have studied atomic  Fermi-Bose mixtures in 
optical lattices in the strong interaction limit, and in the 
presence of an inhomogeneous, or random
 on-site potential. We have derived the effective Hamiltonian describing 
the low temperature physics of the system, and shown  that 
an inhomogeneous potential may be efficiently 
used to control the nature and strength of (boson mediated) 
interactions in the system. Using  a  random potential, one is able to control 
the system in such a way that its physics  corresponds to 
a whole variety of quantum disordered systems: 
Fermi glass, fermionic spin glass, and 
quantum percolation systems.  

We thank  M. Baranov, H.-P. B\"uchler,  
A. Georges, J. Wehr, J. Parrondo and P. Zoller for fruitful discussions.
We acknowledge support from the Deutsche
Forschungsgemeinschaft (SFB 407 and SPP1116), the RTN Cold Quantum Gases, 
ESF PESC BEC2000+, the Alexander von Humboldt Foundation and KBN grant
PBZ-MIN-008/P03/2003 (J.Z.).

\end{document}